\documentclass[twocolumn,showpacs,aps,prl,floatfix,superscriptaddress]{revtex4}
\usepackage{graphicx}
\usepackage{citesort}
\usepackage{epic,eepic}
\usepackage{graphics}
\usepackage{epsfig}
\usepackage{subfigure}
\newcommand{\G}{{\cal G}}

\newcommand{\be}{\begin{equation}}
\newcommand{\ee}{\end{equation}}
\usepackage{amsmath}
\usepackage{amsfonts}
\usepackage{amssymb}
\usepackage{bm}
\usepackage{color}
\usepackage{graphicx}
\begin{document}
\title{A  mesoscopic approach for multi-phase flows in
  nano-corrugated channels}
\author{R. Benzi} \affiliation{Dipartimento di Fisica and INFN,
  Universit\`a di Roma ``Tor Vergata'', Via della Ricerca Scientifica
  1, 00133 Roma, Italy.}
\author{L. Biferale} \affiliation{Dipartimento di Fisica and INFN,
  Universit\`a di Roma ``Tor Vergata'', Via della Ricerca Scientifica
  1, 00133 Roma, Italy.}
\author{M. Sbragaglia} \affiliation{Department of Applied Physics, University of Twente, P.O. Box 217, 7500 AE Enschede, The Netherlands} 
\author{S. Succi} \affiliation{Istituto per le Applicazioni del
  Calcolo CNR, Viale del Policlinico 137, 00161 Roma, Italy.}
\author{F.\ Toschi} \affiliation{Istituto per le Applicazioni del
  Calcolo CNR, Viale del Policlinico 137, 00161 Roma, Italy.}
  \affiliation{ INFN, Sezione di Ferrara,
  via G. Saragat 1, I-44100, Ferrara, Italy.}
\date{\today}

\begin{abstract}
   An approach based on a lattice version of the
   Boltzmann kinetic equation for describing multi-phase flows in nano-
   and micro-corrugated devices is proposed. We specialize it to
   describe the wetting/dewetting transition of fluids in presence of
   nanoscopic grooves etched on the boundaries.  This approach permits
   to retain the essential {\it supra-molecular} details of fluid-solid
   interactions without surrendering -actually boosting- the
   computational efficiency of continuum methods.  The mesoscopic
   method is first validated quantitatively against Molecular Dynamics
   (MD) results of
 {\it Cottin-Bizonne et al.} [Nature Mater. {\bf 2} 237 (2003)]
and then applied to more complex situations which are hardly
   accessible to MD simulations.  The resulting analysis 
   confirms that  surface roughness and capillary effects may conspire to
   promote a counter-intuitive but significant reduction of the flow
   drag with substantial enhancement in the mass flow rates
   and  slip-lengths in the micrometric range for highly
   hydrophobic surfaces. 
\end{abstract}
            
\maketitle
\begin{figure}
\begin{center}
\includegraphics[scale=.6]{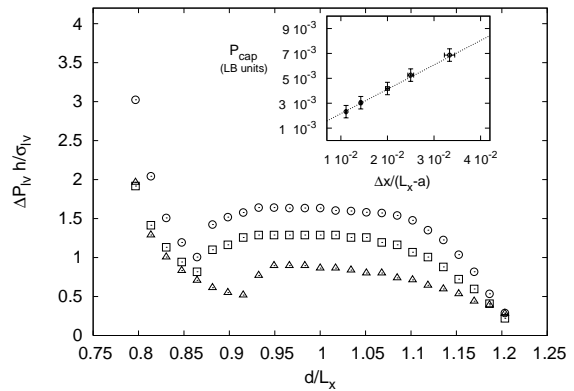}
\caption{Pressure variations at changing $d$ for various roughnesses, $L_x =
  50\Delta x$ ($\bigtriangleup$), $L_{x}=60 \Delta x$ ($\Box$), $L_{x}=80
  \Delta x$ ($\circ$) with $a= 10 \Delta x$.  In the inset, we show
 the relation (\ref{G}). The slope is given by $2
  \sigma_{lv} \cos(\theta)=-0.196 \pm 0.006$ (LB units).  With our
  surface tension $\sigma_{lv}=0.105\pm 0.002$ (LB units), this
  implies a best estimate of $\theta=158^o \pm 6^o$.}
\label{fig:2}
\end{center}
\end{figure}

The motion of fluids at the micro and nanoscale is controlled by the
competition of dissipative effects and pressure drive. 
 The weakness of inertia in the microworld
implies that it is increasingly difficult to push fluids across
micro/nanoconfined geometries, as their surface/volume ratio is made
larger and larger.  
An obvious consequence is that the dynamics of microflows is crucially
affected by the interaction of the fluid with the confining solid
boundaries.  Information on these interactions is usually conveyed
into the formulation of proper boundary conditions  for the fluid flow.
\noindent 
In particular, slippage properties (see \cite{brenner} for a  recet review) 
have been reported in
experiments and in molecular dynamics simulations, depending on the
thermodynamical and wetting properties of the boundary
(contact angle) and on the surface geometry
\cite{Zhu,Ou,Cottin,Cheng,Watanabe,Baudry,Barrat,MD5}.
A fundamental question arises as to whether the fluid really slips
over the surface, or rather the indirect measurements based on
pressure/mass flow rate relations  or
surface force apparatus reflect an
apparent slip arising from surface inhomogeneities or complex
interface with additional physics. Indeed, it has been argued that a
gas layer at the interface would alter the fluid dynamics in the bulk,
leading to a mass flow rate increase even in the presence of pure
no-slip \cite{laugabrenner,thetheway,kwok,benzi,hart}. This
hypothesis is supported by the observation of nanobubbles trapped on
the surface \cite{tyrrel} and by a decreasing
apparent slip length as the fluid is degased \cite{Granick}.\\
\noindent The aim of this article is to discuss the complex effects,
at the hydrodynamical scales, induced by the surface wetting
properties in presence of complex geometries in micro- and
nano-devices.  The results can be summarized in two main
points. First, we provide neat evidence that the physics of the
boundary conditions is quantitatively reproduced by modeling the fluid
at mesoscopic level, by means of a minimal version of the Boltzmann
equation, i.e. the Lattice Boltzmann Equation (LBE)
\cite{chenreview,BSV}.  This result is obtained by
performing a quantitative comparison  of a  ``finite-volume'' dewetting transition against recent Molecular Dynamics
simulations (MD) \cite{Barrat,Barrat2}.  Far from being a mere
technicality, this result opens the way to numerical investigations at
spatial and time scales much larger than those currently available in
most MD simulations.
Second, we extend the MD results by investigating the critical
dependency of the mass flow rate on the degree of roughness at
constant bulk pressure.  The simplest LBE reads as follows
\cite{LBGK}:
\begin{eqnarray}
&f_i(\bm{x}+ {\bm c}_i \Delta t,t+\Delta t)-f_i(\bm{x},t)= \nonumber \\
&-\omega \Delta t [f_i(\bm{x},t) -f_i^{(eq)}(\bm{x},t)]+F_i \Delta t,
\label{1}
\end{eqnarray}
where $f_i({\bm x},t)$
 is the probability of finding a particle at site ${\bm x}$ 
at time $t$, moving along one of the $b$-{\em th} lattice
direction defined by the discrete speed ${\bm c}_i$ with $i=1,\dots,b$
and $\Delta t$ is
the time unit.  The left-hand side of (\ref{1}) stands for molecular
free-streaming, whereas the
right-hand side represents molecular collisions. These are expressed
through a simple relaxation towards local Maxwellian equilibrium
$f_i^{(eq)}$ in a time lapse of the order of $\tau \equiv
\omega^{-1}$.  Finally the term $F_{i}$ represents a volumetric
body-force, which can be tailored to produce highly non-trivial
macroscopic effects, such as phase-transitions.  Non-ideal effects,
leading to two-phase flows, are modeled through a self-consistent force term: 
\be
\label{2} {\bm F} ({\bm x},t)= \G_b \sum_i w_i \psi({\bm x},t) \psi({\bm
  x}+{\bm c}_i \Delta t,t) {\bm c}_i. \ee 
Here, $\psi({\bm x})$ is a
phenomenological pseudo-potential (generalized density),
 $\psi({\bm x},t)=\psi[\rho({\bm x},t)]$, first introduced 
by Shan and Chen \cite{SC}, $w_i$ are normalization weights  and  $\G_b$
tunes the molecule-molecule interaction, i.e. it plays the role of the
normalized {\it inverse temperature}, $\epsilon/KT$, with $\epsilon$ 
the molecular interaction, $K$ The Boltzmann constant and $T$ the system temperature.  Here we choose the standard form $\psi =
\sqrt{\rho_0} \{1-\exp(-\rho/\rho_0)\}$, with the reference density
$\rho_0=1$, in lattice units. \\
In spite of its simplicity, the Shan-Chen approach provides two
crucial ingredients of non-ideal fluid behavior: a non-ideal equation of
state and a non-zero liquid-vapor surface tension, $\sigma_{lv}$.  Both
features are encoded in the expression of the non-ideal momentum flux
tensor $P_{kj}$.  In the hydrodynamic limit, the LBE equations
(\ref{1})-(\ref{2}) can be shown to evolve according to the
Navier-Stokes equations \cite{jfm} with a 
pressure tensor $ \overleftrightarrow{P}$:
\begin{eqnarray}
\label{TENSORE}
& P_{kj}=\left[
   c^{2}_{s}\rho+\frac{1}{2}c^{2}_{s}\G_{b}\psi^{2}+\frac{1}{2}c^{4}_{s}\G_{b}\psi
   \Delta \psi +  \frac{\G_{b} c^{4}_{s}}{4} |{\bm \nabla} \psi|^{2}
   \right] \delta_{kj} \nonumber \\ &-\frac{1}{2}c^{4}_{s}\G_{b} \partial_{k} \psi
\partial_{j}\psi.
  \end{eqnarray}
The equation of state in the bulk is $P=
  c^{2}_{s}\rho+\frac{1}{2}c^{2}_{s}\G_{b}\psi^{2}$ where we recognize
  a non-ideal contribution on top of the ideal equation of state
  $P=\rho c_s^2$ with $c^{2}_{s}$ the sound speed velocity.  This
  equation of state supports a phase-transition at a critical density
  $\rho_c/\rho_0= \ln(2)$, whenever the coupling strength exceeds (in
  magnitude) the critical value $\G_c=-4.0$.  Concerning the velocity
  field, following \cite{jfm}, we set it to be zero at the boundary by
  using bounce-back boundary conditions while, for the case when
  slippage properties are apriori imposed, different boundary
  conditions should be used \cite{karli2,jfm2}.
\begin{figure}[h]
\begin{center}
\includegraphics[scale=.6]{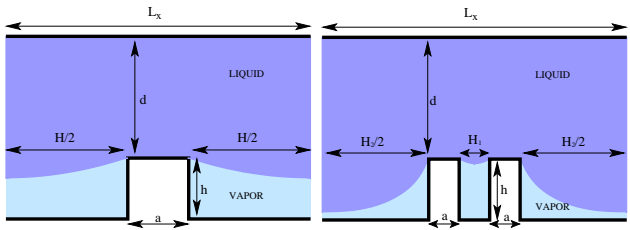}
\caption{Left: homogeneous roughness. A groove with depth $h=33 \Delta x$  and width $H = L_x-a$ (with $a=10 \Delta x$) is introduced on the bottom wall and 
  periodic boundary conditions are assumed along $x$.
  In this configuration, the
  presence of vapor pockets inside the groove changes the
  ``effective'' boundary conditions felt by the bulk fluid, with a net
  decrease of drag when a pressure drop is applied.  Right: channel
  with heterogeneous roughness. Two grooves of width $H_1 = 40 \Delta
  x $ and $H_2 = 70 \Delta x$ are present. The two grooves are filled
  separately at different values in the pressure/density diagram.  The
  lattice spacing corresponds to $\Delta x \sim 0.3 nm$. }
\label{fig:0}
\end{center}
\end{figure}
Next, the density field $\rho({\bm x})$ is assumed to match a given
value $\psi_w = \psi(\rho_w)$, where $\rho_w$ should be regarded as a
free-parameter related to the strength of the fluid/solid
interactions.  In \cite{jfm} we have shown that by imposing the
condition of mechanical equilibrium, $\partial_j P_{kj} =0$, of the
contact line separating the liquid, vapor and solid phases, one can
compute analytically the {\it contact angle}, $\theta(\G_b,\rho_w)$.
Notwithstanding their inherently mesoscopic character, the parameters
$\G_b$ and $\rho_w$ carry no less physical content than their
atomistic counterparts (relative strength of the attractive to
repulsive interactions in MD simulations \cite{MD5}).  In order to
study the wetting/dewetting transition on micro/nano-patterned
surfaces (see Fig. \ref{fig:0}) we have integrated the LBE equation
(\ref{1}) in a $2D$ Lattice using the nine-speed $2DQ9$ model ($b=9$),
one of the most used $2D$-LBE scheme, due to its superior stability
\cite{KARLI,bruce}.  We have used the two geometries described in
Fig.~\ref{fig:0}. All simulations have been performed at fixed
``inverse temperature'' $\G_b=-6.0$ i.e. beyond the critical value
$\G_b=-4.0$. The relaxation parameter is taken as $\tau=0.8$ in
lattice units. The equation of state delivers the corresponding values
of the liquid and gas density, $\rho_l=2.65$ and $\rho_g=0.07$
respectively, while the surface tension is $\sigma_{lv}= 0.105 \pm
0.002 $ in lattice units.  The value of the lattice spacing in
physical units is obtained by matching the physical value of the
liquid-vapor surface tension, with the one measured on the lattice,
via the dimensional relation: $\sigma_{lv}^{phys} = KT /(\Delta x)^2
\sigma_{lv}$. For water/vapor at $T=540^o$, where the density ratio is
close to the values of our simulation, we have $\sigma_{lv}^{phys} =
0.022 N/m$ which yelds $\Delta x \sim 0.3 nm$.  This value is
comparable with the atomic range of the Lennard-Jones potential used
in MD.  Molecular dynamics simulations have recently reported that the
concerted effects of wetting phenomena and nano-corrugations can lead
to a fairly substantial reduction of mechanical drag
\cite{Barrat,Barrat2}.  Specifically, the authors in
\cite{Barrat,Barrat2} consider a nanometric channel flow with a
regular sequence of longitudinal or transverse steps (with respect to
the mean flow) along the solid wall of the channel, and show that,
under suitable thermodynamic and geometric conditions, the presence of
the steps triggers the formation of a gas film in the grooves within
the obstacles.  The liquid can then slide-away over the gas film,
thereby experiencing a significantly reduced mechanical drag.  Such
phenomenon may occur only at a critical pressure drop between liquid
and vapor phase, of the order of the capillary pressure, $P_{cap}$,
given by the estimate: \cite{Barrat2}: \be
\label{G}
P_{cap} = -\frac{2 \sigma_{lv}\cos(\theta)}{L_x-a}.  
\ee 
In Fig.~\ref{fig:1}, we validate the LBE by a direct comparison with
the MD results published in \cite{Barrat,Barrat2} with the same
geometry and with comparable contact angle.  In particular, we
show the Pressure drop between the bulk liquid and vapor phases
$\Delta P_{lv} = P_l - P_v$, at changing the distance $d$ (see Fig.~\ref{fig:0}) at fixed
total mass, i.e. at changing the average density.  As one can clearly
see, the agreement between LBE and MD is quantitative. The plateaux
observed for $\Delta P_{lv}$ in the range $0.9 < d/L_x < 1.05 $
corresponds to the pressure/density values at which the fluid is
invading the corrugation, forming an interface which does not yet
touch the bottom of the groove. This corresponds to the capillary
pressure, $P_{cap}$. Reducing further $d$, i.e. increasing the average
density, a change of concavity for $0.8 < d/L_x < 0.9 $, is observed.
This range corresponds to values such that the interface starts to
touch the bottom of the groove, adjusting its pressure/density in such
a way as to minimize the free energy in the presence of the new
liquid-vapor-solid interface. 
\begin{figure}
\begin{center}
\includegraphics[scale=.6]{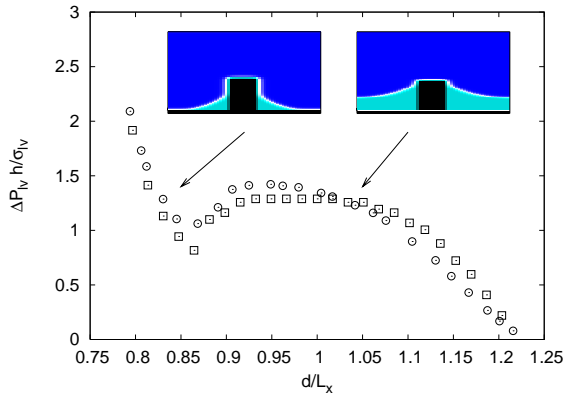}
\caption{Validation of our LBE simulation against MD results
  \cite{Barrat2}. The LBE is done with the same aspect-ratios of MD.
 The dimensionless normalized pressure drop,
  $\Delta P_{lv} h/\sigma_{lv}$ between the two bulk phases, is shown as a
  function of the normalized distance $d/L_{x}$ (see Fig.~\ref{fig:0}).
LBE ($\Box$) results have been obtained
  with a contact angle $\theta = 160^o$, which is consistent with the
  contact angle measured in MD ($\circ$) simulation (see Fig. 2 of
  \cite{Barrat2}). The two insets represent the density configuration
  at the onset of the wetting/dewetting transition (right) and for a
  wetted configuration (left). The plateaux in the pressure curve
  defines the capillary pressure, $P_{cap}$.}
\label{fig:1}
\end{center}
\end{figure}
The agreement of LBE with the capillary
pressure (\ref{G}) is checked in detail in Fig.~\ref{fig:2},
where we report the change of the pressure diagram with a changing
distance $d$, for three different corrugation values, $L_x-a$.  In the
inset of Fig.~\ref{fig:2}, we extract the value of $\sigma_{lv}
\cos(\theta)$ from the the slope of the observed plateaux {\it vs} $ \Delta x/(L_x-a) $. The value of $\cos(\theta)$ is then obtained by estimating the surface
tension, $\sigma_{lv}$, through Laplace's law for a droplet in
equilibrium with its saturated vapor. The agreement of the contact
angle measured in this way, $\theta = 158^o \pm 6^o$, with the
analytical estimate, $\theta = 160^o$, obtained in \cite{jfm} by
imposing the mechanical equilibrium condition of the contact line  is very satisfactory.  \\
The comparison shown in Fig. \ref{fig:1} and Fig. \ref{fig:2}
demonstrates the first result discussed in this paper, namely that the
model introduced in \cite{jfm} captures the correct interplay between
roughness and wetting effects.  \\
Even more complex behavior is observed for heterogeneous
nano-corrugations, with the simplest case shown in the right panel of
Fig.~\ref{fig:0}. In this case, one has two characteristic groove
sizes, $H_1$ and $H_2$, and hence two corresponding critical capillary
pressures. The pressure/density diagram for this heterogenous
corrugation is shown in Fig. \ref{fig:3}, where the two plateaux
corresponding to the two capillary pressures coexist. This device may
be considered a ``smart'' two-state surface, whose wetting properties
and mass throughput (under the application of a pressure
gradient ) may be tuned by  changing the bulk pressure. \\
The dynamical response of the micro-channel is investigated by
applying a constant pressure gradient. In Fig.~\ref{fig:4}, we show
the mass flow rate as a function of the degree of roughness, $a/(L_x-a)$,
for a fixed groove depth $h$ and at a given normalized pressure drop,
$\Delta P_{lv} h /\sigma_{lv} = 0.75 $.  The main result here is the
presence of a  transition at a critical
roughness, $a/(L_x-a)$, where the mass flow rate starts to increase
with respect to the perfect wetting situations reaching as much as
$100\%$ gain.  The strong dynamical effect of the gas-layer can be 
quantified in more
detail by inspecting the velocity and momentum profiles along the
vertical direction without the vapor layer (fully-wetted
configuration) and with a thin vapor layer starting to accumulate
close to the bottom of the groove.  
This is shown in the inset of Fig.~\ref{fig:4}
where the {\it local } slip length is also depicted by extrapolation
of the bulk profile inside the wall. As soon as a vapor layer is
formed, the {\it local} slip-length ramps-up, reaching values of the
order of the channel height $ 45 \Delta x \sim 15 nm$.  Even larger
values can be measured close to the dewetting transition, where a
well-developed vapor layer is
formed inside the groove (the so-called {\it super-hydrophobic regime}).  \\
\noindent 

Summarizing, we have shown that an extension of lattice Boltzmann 
equation for non-ideal fluids,
can {\it quantitatively} account for the concerted effects between
wetting phenomena and geometrical irregularities.  
In particular, the presence of nano/micro-irregularities in the flow
geometries leads to sizeable effects with respect to the {\it infinite
  volume} liquid-gas transitions, as well as to a significant
reduction of mechanical drag on the flowing fluid.  The consequence of
our results is twofold: from a theoretical perspective, it indicates
that drag-reduction via geometry-induced wetting transitions is a
non-specific phenomenon.  On the practical side, the present LBE
approach offers the opportunity to perform very efficient numerical
simulations of complex micro/nanofluidic phenomena at scales of direct
experimental relevance, which are hardly accessible to atomistic
simulations.  This will open the way to the systematic optimization of
microfluidic devices via computer simulation.
\noindent 
Useful discussions with J.-L. Barrat, X. Shan and S. Troian are kindly acknowledged. 
\begin{figure}
\begin{center}
\includegraphics[scale=.6]{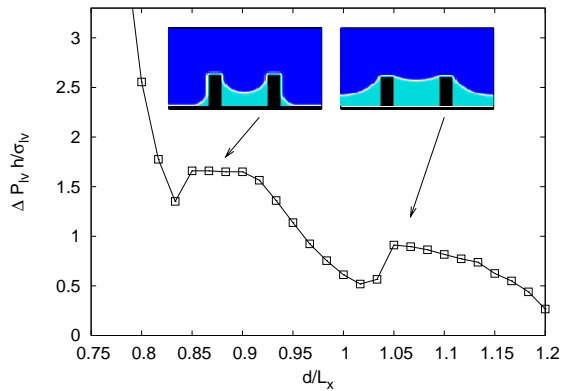}
\caption{Pressure variations at changing $d$, with fixed total mass
  for the case of heterogeneous roughness (see right panel of
  Fig.~\ref{fig:0}).  The two plateaux correspond to the case in which
  the liquid is starting to invade the widest groove (right) and when
  it is invading also the thinnest (left).}
\label{fig:3}
\end{center}
\end{figure}
\begin{figure}
\begin{center}
\includegraphics[scale=.6]{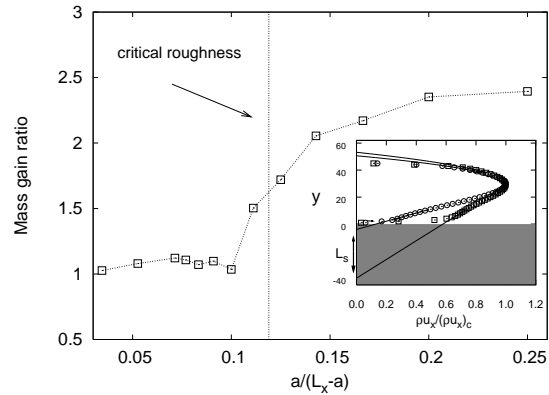}
\caption{Mass flow rate normalized to fully wetted case {\it vs } the effective roughness
  $a/(L_{x}-a)$. The bulk pressure is fixed to be $\Delta P_{lv} =
  0.75 \sigma_{lv}/h$.  A critical
  roughness (vertical line) is given by the estimate of the capillary pressure,
  $0.75\frac{a}{2 h \cos(\theta)}=0.119$ for $a=10
  \Delta x, h=33 \Delta x$, $\theta=158^o$. 
  Inset: Vertical momentum profile 
  for a wetted ($\circ$),  and almost
  dewetted configuration ($\Box$). The geometry is fixed to the one of
  the left panel of Fig.~\ref{fig:0} with parameters $h=14 \Delta x$,
  $a=14 \Delta x$, $L_{y}=45 \Delta x$, $L_{x}=90 \Delta x$. Both
  momentum profiles are shown for $x/L_{x}=0.1$ and normalized
  with their center channel values.  The straight lines correspond to
  extrapolations of the profiles inside the boundaries, i.e. the
  standard way to define a slip-length. }
\label{fig:4}
\end{center}
\end{figure}

\end{document}